\documentclass[%
 reprint,
 amsmath,amssymb,
 aps,
 prb,
]{revtex4-1}
\usepackage{xcolor}

\usepackage{graphicx}
\usepackage{dcolumn}
\usepackage{bm}
\usepackage{bbold}
\usepackage{appendix}
\usepackage{hyperref}

\usepackage[normalem]{ulem}

\usepackage{caption}
\usepackage{subcaption}

\usepackage{algorithm}
\usepackage{algpseudocode}

\newcommand{\vthree}[1]{{#1}}

\newcommand{\vone}[1]{{}}
\newcommand{\vtwo}[1]{{#1}}

\begin{document}

\title{The quantum Gaussian process state: A kernel-inspired state with quantum support data}

\author{Yannic Rath}
\email{yannic.rath@kcl.ac.uk}
\affiliation{Department of Physics, King’s College London, Strand, London WC2R 2LS, United Kingdom}%

\author{George H. Booth}
\email{george.booth@kcl.ac.uk}
\affiliation{Department of Physics, King’s College London, Strand, London WC2R 2LS, United Kingdom}%

\date{\today}

\begin{abstract}
We introduce the quantum Gaussian process state, motivated via a statistical inference for the wave function supported by a data set of unentangled product states.
We show that this condenses down to a compact and expressive parametric form, with a variational flexibility shown to be competitive or surpassing established alternatives.
The connections of the state to its roots as a Bayesian inference machine as well as matrix product states, also allow for efficient deterministic training of global states from small training data with enhanced generalization, including on application to frustrated spin physics.
\end{abstract}
\maketitle
\makeatletter
\section{Introduction}
From the simulation of quantum phase transitions to modeling the qubits of quantum computers, accurately describing the quantum state of a many-body system is central to many fields.
With the exact quantum state suffering from an exponential curse of dimensionality due to its superposition of all `classical' configurations of particles, the field is driven by computationally tractable and accurate representations of these states.
Traditional parameteric models are motivated by spanning certain physical features of a correlated state, such as the Jastrow \cite{PhysRev.98.1479} or correlator product state (CPS) \cite{ Mezzacapo_2009, PhysRevB.80.245116, PhysRevB.85.045103, PhysRevB.85.060402}.
However while these are polynomially compact representations, it is not straightforward to systematically improve the expressiveness of these models.
Recently, the field of machine learning (ML) has brought about a new perspective on wave function models.
By virtue of these states inheriting model properties as universal approximators, they have allowed systematic control over their flexibility and expressiveness to be a central tenet in their construction.
This enables an automatic construction of correlated many-body features of the state, analogous to the extraction of characterizing features in e.g. object selection in image recognition.
In many-body physics, these models have overwhelmingly taken the form of neural quantum states (NQS) of various architectures, including feed-forward, recurrent, convolutional and restricted Boltzmann machines \cite{Carleo602, Gao_2017, PhysRevB.96.205152, PhysRevB.97.035116, carleo2018constructing, PhysRevLett.121.167204, PhysRevB.98.104426, PhysRevB.100.125124, PhysRevB.101.195141, Hermann_2020, PhysRevResearch.2.033429, Kessler_2021, PhysRevLett.124.097201, Westerhout_2020, PhysRevResearch.2.012039, PhysRevE.101.053301, 10.21468/SciPostPhys.10.6.147, Nomura_2021, PhysRevResearch.2.033075, nomura2021diractype, sharir2021neural, vieijra2021manybody, roth2021group, pei2021compact, Clark_2018, PhysRevB.100.125131, PhysRevX.8.011006, park2021expressive, chen2021neural}.

This family of systematically improvable quantum states was recently enlarged with an alternate paradigm of ML-inspired model construction.
`Kernel methods' encompass Gaussian process regression, kernel ridge regression and support vector machines, and are characterized by linear models after projection into a high-dimensional non-linear feature space \cite{Rasmussen_book}.
In these approaches, data points explicitly support the model definition, along with a kernel function, which represents the covariance between different configurations in the prior distribution, and defines an inner product between configurations in this high-dimensional space of features.
By writing a quantum state as a probablistic Gaussian process and defining an appropriate kernel function between many-body configurations, the `Gaussian process state' (GPS) was introduced in Ref.~\onlinecite{PhysRevX.10.041026} as a systematically improvable quantum state.
This has a simple functional form, which benefited from an underlying Bayesian interpretation for a sparse learning of the state, as well as a rigorous mathematical underpinning of regularization and generalization characteristics \cite{doi:10.1063/5.0024570}.

In this work, we show how the data which defines this model can be provided as unentangled product states rather than classical configurations, which results in a sharp increase in the flexibility and rate of convergence of the model.
This provides a systematically improvable state with a simple functional form, \vtwo{which is fully defined by data on a continuously varying manifold, rather than requiring a set of discretely varying classical data configurations.
Furthermore,} desirable characteristics are inherited from its root as a kernel method, including a Bayesian interpretation, well defined procedures for regularization, and rigorous mathematical underpinning of generalization properties and optimization strategies.
We will also show how this state can also be viewed from a tensor network or neural network perspective.
The confluence of these multiple perspectives on this state -- as a matrix product state, neural quantum state and a Bayesian probablistic model -- enable a number of opportunities for this unique parameterization, which we will demonstrate for a series frustrated and unfrustrated magnetic lattice models.

\section{From classical to quantum data in Gaussian Process States}
We represent the original GPS as an exponential of the mean of a Gaussian process, as
\begin{equation}
    \Psi(\mathbf{x}) = \exp \left(\sum_{\{\mathbf{x}'\}} w_{\mathbf{x}'} k(\mathbf{x}, \mathbf{x}') \right), \label{eq:ClassGPS}
\end{equation}
where $\Psi(\mathbf{x})$ is the predicted probability amplitude on many-body test configuration $\mathbf{x}$, with the model supported by a set of $M$ many-body `classical' configurations, $\{ \mathbf{x}' \}$. \vthree{These many-body configurations, $\mathbf{x}$ and $\mathbf{x}'$, define specific local Fock states for each degree of freedom in the system.}
The presence of these `support' configurations in the definition contrasts with other ML-inspired ansatz, \vthree{as this support set of configurations explicitly enter the state definition,} resulting in an directly data-driven approach \cite{PhysRevX.10.041026}.
The weights of these support configurations are given by $w_{\mathbf{x}'}$, with the kernel function between these support and test configurations given by $k(\mathbf{x}, \mathbf{x}')$.
This kernel function characterizes the similarity between test and support configurations, by implicitly summing the total number of coincident occupations between the two configurations over all plaquettes of sites of any possible shape and size \cite{PhysRevX.10.041026}.
This efficient resumming of all correlated features into a compact kernel function \vthree{(a manifestation of the `kernel trick' in machine learning terminology)} ensures that there is no limit on the modeled correlation range or rank, even with a single support configuration, and the state can be made systematically more complete by increasing the set of support configurations.
\vthree{The definition of the GPS thus depends both on continuous parameters (the weights $w_{\mathbf{x}'}$ for each support configuration, as well as a small number of additional hyperparameters defining the kernel function) as well as discrete variables giving the many-body support configurations selected from the full Hilbert space basis.}

Assuming a discrete many-body system of $L$ sites with a $D$-dimensional local Hilbert space on each site, the (non-symmetrized) GPS kernel which quantifies the similarity of configurations $\mathbf{x}$ and $\mathbf{x}'$ over all plaquettes is found as
\begin{equation}
    k(\mathbf{x}, \mathbf{x}') =  \prod_{i = 1}^L  e^{- (1 -\delta_{\mathbf{x}_i, \mathbf{x}'_i})/f(i)}. \label{eq:classkernel}
\end{equation}
The delta function $\delta_{\mathbf{x}_i, \mathbf{x}'_i}$ evaluates to one if the local Fock state of site $i$ is the same in both configurations (i.e. $\mathbf{x}_i=\mathbf{x}'_i$), and zero otherwise.
A distance-dependent discrete function characterized by a small number of hyperparameters, $f(i)$, is chosen to provide additional flexibility, and modifies the fit to preferentially weight a desired rank or range of implicit plaquettes of correlated features.
These hyperparameters can be variationally optimized, or fit to data in a principled Bayesian approach via maximization of the marginal likelihood of the model \cite{doi:10.1063/5.0024570}.
A drawback of the GPS formulation is the requirement to define the set of discrete support configurations over the same computational lattice, requiring a combined discrete-continuous optimization of the state.
Previously, this was solved via a Bayesian supervised learning of the state \cite{Tipping2000, Tipping2003, Tipping_2004}, again using the marginal likelihood and a sparse prior on candidate support configurations in order to select a compact set of configurations \cite{doi:10.1063/5.0024570}.
This could then be combined with a variational optimization of the weights and/or hyperparameters of the model, if the desired state was not known in advance \cite{PhysRevX.10.041026}.

However, in this work we generalize the state and avoid the requirement to \vthree{select a discrete set of} support configurations, by instead defining the support set as {\it unentangled product states}\vthree{, $\phi_{x'} = \bigotimes_{i=1}^L \phi^{(i)}_{x'}$} resulting in a fully parametric and continuous model.
\vthree{This allows us to simultaneously and variationally optimize all parameters characterizing the support configurations which turns the problem of finding the best representation of a (typically unknown) target state into a fully continuous optimization problem.}
For a spin system ($D=2$), \vthree{such} support states of the model are \vthree{chosen to be} (unnormalized) local superpositions of the ${\hat S_z}$ eigenstates on each site, as
\begin{equation}
    \phi_{x'} = \bigotimes_{i=1}^L \left( \alpha_{\uparrow, x', i} |\uparrow_i \rangle + \alpha_{\downarrow, x', i} |\downarrow_i \rangle \right).
\end{equation}
The use of $x'$ is now simply a label for the `quantum' support point of the model, rather than a discrete \vthree{`classical' many-body} configuration, \vthree{and therefore we now refer to it as a scalar quantity}.
The kernel \vthree{is now modified, by} replacing $\delta_{\mathbf{x}_i,\mathbf{x'}_i}$ \vthree{in Eq.~\ref{eq:classkernel} (a `classical' overlap of the configurations) by the continuous coefficient of the training product states $\alpha_{\mathbf{x}_i,x',i}$. These values define the component of the state $\phi^{(i)}_{x'}$ on site $i$, and therefore the overall} overlap of a test configuration with the support point.
\vthree{This approach directly defines the kernel function with respect to the support point labeled by $x'$ according to
\begin{equation}
    k_{x'}(\mathbf{x}) = \prod_{i = 1}^L e^{- (1 -\alpha_{\mathbf{x}_i, x', i})/f(i)},
\end{equation}
specified by the coefficients $\alpha$.}

\vthree{By re-expressing the functional form with respect to the parameters $\epsilon_{\mathbf{x}_i, x', i} = w_{x'}^{1/L} e^{- (1 -\alpha_{\mathbf{x}_i, x', i})/f(i)}$}, the weights \vthree{and hyperparameters of the kernel} can be further subsumed into the definition of the ansatz as a form of `kernel learning'
, giving the final model a \vthree{simple} functional form of
\begin{equation}
    \label{eq:QGPS_def}
    \Psi(\mathbf{x}) = \exp \left(\sum_{x'=1}^M \prod_{i=1}^L \epsilon_{\mathbf{x}_i, x', i} \right)\vthree{.}
\end{equation}
\vthree{This model} is entirely parameterized by the $ D \times M \times L$ tensor of complex-valued variational parameters, $\epsilon_{\mathbf{x}_i, x', i}$.
We define this as a `quantum' Gaussian process state (qGPS), to distinguish it from the use of simple spin configurations as support points of the previous model \vthree{(which we continue to denote as the GPS)}.
We denote the parameter $M$ as the support dimension of the model, which was previously the dimension of the support configuration set, and is the only hyperparameter which needs to be chosen to fully specify the parametric class spanned by the qGPS.
For a fixed $M$, any GPS can be spanned by an equivalent qGPS, with the qGPS having significant additional variational flexibility.
Furthermore, the qGPS is systematically improvable to exactness as $M$ is increased.

Viewing the qGPS as a simple parameteric variational ansatz, it is an exponentiated multi-linear estimator for each configurational amplitude.
This exponential form ensures product separability of the weighted features in the model prediction of each amplitude, and therefore size extensivity of resulting energies.
It can be expanded as a power series to express the qGPS state a \vthree{weighted sum over all possible products from the set of $M$ unentangled states, with each term in the sum introducing additional entanglement in the qGPS. The state can therefore be viewed} as an infinite \vthree{(albeit parameterized)} linear combination of \vthree{non-orthogonal} product states \vthree{, i.e. we can represent the qGPS amplitudes as
\begin{equation}
    \Psi(\mathbf{x}) = \sum_{l=0}^\infty \frac{\left(\sum_{x'=1}^M \prod_{i=1}^L \epsilon_{\mathbf{x}_i, x', i} \right)^l}{l!}.
\end{equation}
We therefore expect this state to be able to express volume-law scaling entanglement in a similar way as neural quantum states \cite{PhysRevX.7.021021, https://doi.org/10.48550/arxiv.2203.00020}, although leave detailed theoretical and numerical studies of this entanglement scaling of the qGPS for future investigations.}
This perspective also allows connection to the field of matrix product states (MPS), where the qGPS can be considered as a exponentiated MPS where all the individual site matrices are constrained to be diagonal (and hence commutative and invariant to site ordering or lattice dimensionality) \cite{ORUS2014117,SCHOLLWOCK201196}.
This connection inspires a deterministic DMRG-style sweep algorithm for supervised learning with this state, which will be described in section~\ref{sec:BayesianOpt}.

While it has been known in the ML community that Gaussian processes can \vthree{in general} map to infinitely wide neural networks \cite{Neal94}, the qGPS state can, for $D=2$, be exactly and constructively mapped to a four-layer deep feed-forward neural network, with specific activation functions and constraints on the connectivity between layers (see section \ref{sec:nn_representation_qGPS} of the \vthree{appendix} for details).
Relating this back to the construction of the original kernel function as a sum over plaquette occupations provides a path to a rigorous and physically motivated architecture of a NQS.
However, a further powerful perspective on this state relies on returning to a Bayesian view, which provides new tools to tackle robust and tractable optimization, regularization and generalization of the resulting state \cite{kochkov2018variational, kochkov2021learning}.
These are fundamental to the practical applicability of the qGPS, and have been found to be key bottlenecks in the applicability of other highly flexible NQS states to challenging many-body problems (as opposed to their variational flexibility) \cite{10.21468/SciPostPhys.10.6.147, Westerhout_2020}\vthree{, which is also a context we return to in Sec.~\ref{sec:BayesianOpt}}.

\section{Variational qGPS}
\begin{figure}
    \centering
    \includegraphics[width=\columnwidth]{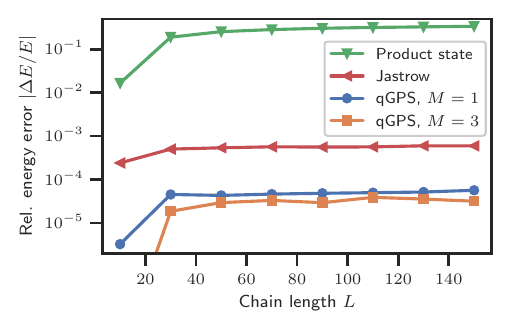}
    \caption{Relative energy error for the ground state of the unfrustrated 1D Heisenberg model as chain length increases, compared to exact DMRG results \cite{fishman2020itensor}.
    Shown are qGPS results with support dimensions $M=1$ and $M=3$, as well as the variational energy of a single projectively symmetrized product state, and the Jastrow ansatz.
    }
    \label{fig:1d_heisenberg}
\end{figure}

We first consider the overall expressibility of the qGPS ansatz and single-parameter improvability via variational Monte Carlo optimization, as implemented in the NetKet package \vthree{\cite{CARLEO2019100311, https://doi.org/10.48550/arxiv.2112.10526}}, with further technical details on the state symmetrization and optimization details provided in sections \ref{sec:qGPS_symmetrization} and \ref{sec:VMC_details} of the \vthree{appendix} \cite{CARLEO2019100311,PhysRevB.64.024512,PhysRevResearch.2.023232}.
We optimize the ground state of the $J_1$-$J_2$ Heisenberg model of $L$ spins, given by
\begin{equation}
    \hat{H} = J_1 \sum_{\langle i, j\rangle} \hat{\mathbf{S}}_i \cdot \hat{\mathbf{S}}_j + J_2 \sum_{\langle \langle i, j \rangle \rangle} \hat{\mathbf{S}}_i \cdot \hat{\mathbf{S}}_j,
\end{equation}
where $J_1$ denotes nearest neighbor coupling, and $J_2$ is the frustration-inducing next-nearest-neighbor interaction.
Numerically exact benchmarks are only available at sign-problem-free unfrustrated points in the phase diagram where $J_2=0$ or 1D, with general frustrated systems still an open problem of significant interest \cite{Mezzacapo_2009, PhysRevResearch.2.033075, Westerhout_2020, PhysRevB.88.060402, PhysRevB.100.125131, Nomura_2021, nomura2021diractype}.
Figure~\ref{fig:1d_heisenberg} considers the 1D $J_2/J_1=0$ model, where the Marshall sign rule is imposed to constrain the exact sign structure \cite{doi:10.1098/rspa.1955.0200}.
This ensures that the qGPS has only to fit the magnitude of the configurational amplitudes.
We find \vthree{that this fitting} is possible to essentially arbitrarily high accuracy, with excellent results even when the model is defined with respect to only a single quantum support product state.
In this limit, the number of parameters in the model is the same as a single unentangled product state, but with an energy error many orders of magnitude less than this simple (symmetrized) state definition. \vtwo{This is due to the exponential form generating an infinite linear combination of product states which involve all different powers of the individual site occupations, and giving rise to significant entanglement which is missing in a single product state.}
Furthermore, we find that relative energy errors are remarkably consistent as the chain length increases to 150 sites, providing appropriately extensive energies with a fixed support dimension.

\begin{figure}
    \centering
    \includegraphics[width=\columnwidth]{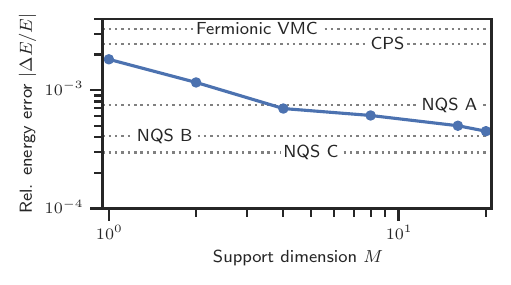}
    \caption{Relative energy errors for the qGPS for the $10 \times 10$ Heisenberg model (defined by $200 \times M$ complex variational parameters) as the support dimension ($M$) is increased, compared to the unbiased results from the stochastic series expansion \cite{PhysRevB.56.11678}.
    Also shown are energy errors from various other ansatzes in the literature (grey dotted horizontal lines).
    From top to bottom the comparison results correspond to a fermionic Gutzwiller-projected mean field description \cite{PhysRevB.88.060402,PhysRevB.100.125124}, CPS \cite{Mezzacapo_2009}, and multiple NQS values from the literature with different architectures (NQS A (3200 complex var. parameters) \cite{Carleo602}, NQS B (5145 real var. parameters) \cite{PhysRevResearch.2.033075}, NQS C (3838 complex var. parameters) \cite{PhysRevB.100.125124}).}
    \label{fig:2d_heisenberg}
\end{figure}

We demonstrate the systematic improvement in expressibility of the state with increasing support dimension of the qGPS in the results of Fig.~\ref{fig:2d_heisenberg} for a $10 \times 10$ 2D lattice.
Here, a support dimension of one gives a relative energy error of $\varepsilon \sim 2 \times 10^{-3}$ (improving on the Gutzwiller projected mean-field \cite{PhysRevB.88.060402,PhysRevB.100.125124} and CPS descriptions \cite{Mezzacapo_2009}) which can be decreased to $\sim 4 \times 10^{-4}$ ($M=20$).
This accuracy is competitive or surpasses the accuracy of recent NQS results for this model across different network architectures \cite{Carleo602,PhysRevResearch.2.033075,PhysRevB.100.125124}.

\begin{figure}
    \centering
    \includegraphics[width=\columnwidth]{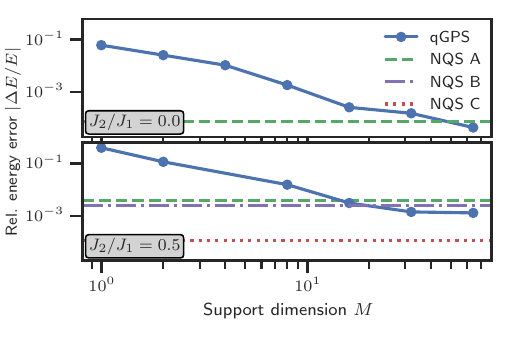}
    \caption{Relative energy errors of the $6 \times 6$ $J_1-J_2$ Heisenberg model as the projectively symmetrized qGPS support dimension ($M$) is increased compared to exact diagonalization \cite{refId0}.
    Results shown for $J_2/J_1=0$ (top), as well as the highly frustrated case $J_2/J_1=0.5$ (bottom), without the Marshall sign rule imposed. Literature NQS values are obtained from Refs.~\onlinecite{PhysRevB.100.125124} (green), \onlinecite{chen2021neural} (purple) and \onlinecite{Nomura_2021} (red).}
    \label{fig:2d_j1_j2}
\end{figure}

It has become clear in a number of studies that learning the sign information for NQS architecture is far more challenging than the amplitude information \cite{10.21468/SciPostPhys.10.6.147, Westerhout_2020, PhysRevResearch.2.033075, chen2021neural}, and we now address this for a qGPS description.
The lack of numerically exact methods for frustrated 2D lattices means that we study the $6 \times 6$ lattice, where exact diagonalization is still feasible, and has become the {\it de facto} testbed for this problem \cite{10.21468/SciPostPhys.10.6.147, PhysRevB.100.125124}.
In Fig.~\ref{fig:2d_j1_j2}, we consider $J_2/J_1=0$, but where the Marshall sign rule (MSR) is no longer imposed and must be learned by the qGPS, and the highly-frustrated $J_2/J_1=0.5$ point where there is no simple rule for the sign of the state.
In these results, we apply a projective symmetrization of the qGPS, rather than symmetrizing the kernel which we find to result in a more robust optimization in cases where sign information is also required \cite{Nomura_2021} (see section \vthree{\ref{sec:qGPS_symmetrization}} of the \vthree{appendix} for more details).
For the $J_2/J_1=0$ results, the errors at low $M$ are larger than when the MSR is imposed {\it a priori} in Fig.~\ref{fig:2d_heisenberg} (notwithstanding the difference in lattice size and symmetrization).
However for larger $M$, the results are comparable, and for $M=64$ even surpass the best state-of-the-art accuracy for variational descriptions in the literature of this unfrustrated point, showing that the MSR can be learned within the ansatz without additional difficulty.

A key research frontier for ML-inspired variational states concerns the ability to learn non-trivial sign structure, required for Fermionic states or at points of significant magnetic frustration, shown in the lower plot of Fig.~\ref{fig:2d_j1_j2}.
For NQS representations, a relative energy error of $\sim 4 \times 10^{-3}$ for the $6 \times 6$ lattice at this frustrated $J_2/J_1=0.5$ point has been suggested as a `universal bottleneck', regardless of architecture flexibility \cite{10.21468/SciPostPhys.10.6.147}, although this was recently improved via imposing symmetries of the \vthree{state} projectively \cite{Nomura_2021} and by introduction of a specific optimization protocol \cite{chen2021neural}.
We find that the qGPS is able to reduce the error from this standard NQS accuracy at $M=32$ by a factor of roughly three, however further significant improvements on increasing $M$ do not materialize.
This suggests that despite obtaining state-of-the-art results from the simple qGPS model with a single improvable parameter, a similar bottleneck exists.
In Ref.~\onlinecite{Westerhout_2020} it was recently made clear that this was not a result of insufficient variational flexibility, but rather the optimization.
More specifically, it was the ability of properties of the model from a small selection of `training' configurations as chosen in VMC, to appropriately generalize across the Hilbert space in order to optimize global properties of the state \vthree{such as the sign structure. 
From a ML supervised learning perspective, this can alternatively be seen as an issue of appropriate regularization, to avoid overfitting in the optimization based on a small configurational sample.}

\vthree{The perspective of supervised learning for a given quantum state is closely related to the challenges of optimization of an ansatz in the context of VMC, where the state is unknown in advance.
In both cases it is required to find the best possible description of the target state, only based on the wave function information from a small fraction of the full Hilbert space of configurations.
Moreover, it is also possible to formulate an optimization scheme for VMC which is based on iterative supervised learning of states \cite{kochkov2018variational, kochkov2021learning}.
Analyzing the ability to learn a general model of the target state from data might thus also be helpful in order to understand and improve optimization techniques in the context of VMC.
}

\section{Bayesian supervised learning with qGPS} \label{sec:BayesianOpt}
It is this issue for which the qGPS may provide a route forwards, due to a combination of its rigorous Bayesian perspective for principled regularization, as well as its connections to MPS.
We consider supervised learning of a given state from a small set of configurational amplitude training data, which has previously been clearly shown to highlight the general problem of NQS state optimization in frustrated systems \cite{Westerhout_2020}.
\vthree{In order to define a Bayesian learning scheme helping to compress a state given by a limited set of data, we can leverage the connection of the qGPS to its roots as an exponentiated mean of a Gaussian process by casting the qGPS model back into a form where the weights and the kernel function of the model are explicitly exposed, allowing standard techniques of Bayesian inference and learning to be defined. In particular, using a set of parameters $\epsilon_{\mathbf{x}_i, x', i}$, we can represent the kernel-symmetrized qGPS in a form equivalent to Eq.~\ref{eq:ClassGPS}, as
\begin{equation}
    \label{eq:kernel_model}
    \Psi(\mathbf{x}) = \exp \left(\sum_{x'=1}^{M} \sum_{d} \epsilon_{d, x', I} \, \tilde{k}^{(I)}_{x', d}(\mathbf{x}) \right),
\end{equation}
with $d$ labelling all possible local basis states (i.e. $d \in \{\uparrow, \downarrow\}$ in the spin models considered here) of a single chosen reference site, $I$.
The weights of the exponentiated kernel model, $\epsilon_{d, x', I}$, are therefore the variational parameters associated with this arbitrarily chosen reference site, and those variational parameters not associated with the reference site $I$, define the kernel function, which defines the coupling of the reference site with the rest of the system, as
\begin{equation}
    \label{eq:quantum_kernel_definition}
    \tilde{k}^{(I)}_{x', d}(\mathbf{x}) = \sum_{\mathcal{S}} \delta_{\mathcal{S}[\mathbf{x}]_I, d} \cdot \prod_{i \neq I}^L \epsilon_{\mathcal{S}[\mathbf{x}]_i, x', i}.
\end{equation}
This definition includes an optional sum over any discrete symmetry operations we want to impose, $\mathcal{S}$, giving rise to the kernel-symmetrized qGPS model (see section \ref{sec:qGPS_symmetrization} of the appendix for details on the symmetrization of the qGPS).

Because the log-amplitudes of the state are linear in the effective weights of the model for a given reference site (i.e. the parameters $\epsilon_{d, x', I}$) -- a key property of kernel methods in ML -- it is possible to apply standard methodology of kernel ML to the task of inferring the weights from given wavefunction data in a controlled fashion with principled regularization \cite{doi:10.1063/5.0024570}.
By sweeping the choice of the reference site, $I$, iteratively across the lattice and correspondingly updating the values $\epsilon_{d, x', I}$, the qGPS representation can then be iteratively and deterministically learnt from a given training data set.

Importantly, we can define a rigorous statistical model to introduce appropriate regularization of the learning procedure at each reference site $I$. This will allow for statistically rigorous inference of the per-site variational parameters, while also ensuring effective generalization of these parameters to describe amplitudes outside the known training data.
This is achieved via a standard Bayesian modeling assumptions, in which we calculate a \textit{posterior} probability distribution (whose mean provides the best estimate for the regularized reference site weights) via application of Bayes' theorem. This combines a Gaussian model for the \textit{likelihood} of the log-wavefunction data with respect to the model, with another Gaussian distribution for the \textit{prior} probability distribution for the weights in the absence of data.
}
The maximization of the resulting posterior distribution then gives the most probable model parameters based on the given training data (accounting for the chosen prior and likelihood).
Furthermore, \vthree{we can follow a standard \textit{type-II maximum likelihood} scheme in order to find suitable hyperparameters defining the prior and the likelihood distributions}
by maximizing the marginal likelihood, obtained by marginalizing the posterior over all possible models
for the weights of site $I$.

\vthree{This Bayesian inference approach used at each step of the iterative sweeping to find the optimal values of $\epsilon_{d, x', I}$ for a given data set is analogous to the approach developed for supervised learning with the classical GPS, which is presented in detail in Ref.~\onlinecite{doi:10.1063/5.0024570}.
Following that scheme and denoting the logarithm of the target wavefunction amplitude for a many-body configuration $\mathbf{x}$ (i.e. a training data point) as $\phi_{\mathbf{x}}$, we model the likelihood of the log-wavefunction amplitude of the a qGPS model as a normal distribution, with mean given by the output of the kernel model,
\begin{equation}
    f_{\mathbf{x}} = \sum_{x'=1}^{M} \sum_{d} \epsilon_{d, x', I} \, \tilde{k}^{(I)}_{x', d}(\mathbf{x}).
\end{equation}
In contrast to the works of Ref.~\onlinecite{doi:10.1063/5.0024570,PhysRevX.10.041026}, here the parameters are in general complex-valued, requiring appropriate modifications \cite{halliwell2015complex, RJ-2015-006}.
This is achieved by assuming that the real and imaginary part of the log-amplitudes are independent, normally distributed, real random variables, with the same variance, given by the training-configuration-specific parameter $\sigma^2_{\mathbf{x}}/2$.
The likelihood for a single log wavefunction amplitude, $\phi_{\mathbf{x}}$, thus takes the form of a complex normal distribution with mean $f_{\mathbf{x}}$, variance $\sigma^2_{\mathbf{x}}$, and a vanishing pseudo-variance.
The likelihood probability density for each training configuration is therefore defined as
\begin{equation}
    p(\phi_{\mathbf{x}} | \bm{\epsilon}^{(I)}, \sigma^2_{\mathbf{x}}) = (\pi \sigma^2_{\mathbf{x}})^{-1} \exp \left( -\frac{|\phi_{\mathbf{x}}-f_{\mathbf{x}}|^2}{\sigma^2_{\mathbf{x}}} \right),
    \label{eq:likelihood}
\end{equation}
where $\bm{\epsilon}^{(I)}$ denotes the vector of the weights, i.e. a flattened vector of the parameters $\epsilon_{d, x', I}$ associated with the reference site $I$.
If multiple data points are contained in a data set, the likelihood for the data is obtained by taking a product of the likelihoods across the different data points.

Again following the approach introduced for the classical GPS \cite{PhysRevX.10.041026,doi:10.1063/5.0024570}, it is important to enforce that the model to have an approximately constant, configuration-independent variance in the prediction of the actual amplitudes ($\tilde{\sigma}^2$) rather than in the likelihood distribution for the log-amplitudes. This can be achieved via ensuring that each training configuration has its own specific choice of variance in the likelihood for the log-amplitudes, $\sigma^2_{\mathbf{x}}$,
as featured in Eq.~\ref{eq:likelihood},given by
\begin{equation}
    \sigma^2_{\mathbf{x}} = \ln{\left ( 1 + \frac{\tilde{\sigma}^2}{|\Psi(\mathbf{x})|^2} \right)}.
    \label{eq:sigma_def}
\end{equation}

In addition to the likelihood, a further model is introduced which defines the prior distribution for the weights of the qGPS.
This forms the expected distribution of the weight parameters prior to accounting for any specific training data.
We follow a standard approach of modeling the prior as a product of Gaussian distributions centered at zero.
Their variance is controlled by an additional hyperparameter $\alpha^{(I)}$ which defines the inverse variance of the Gaussian prior, $p(\bm{\epsilon}^{(I)} | \alpha^{(I)})$, for the current weights, and is allowed to be different between sites.
The extension to complex-valued variables is again achieved by using a complex Gaussian distribution with a real-valued variance, and vanishing pseudo-variance.

By application of Bayes' theorem, the desired posterior probability distribution for the weights can then be inferred based on the likelihood over all the training data, and the prior distribution for the weights.
Applying Bayes' theorem, the posterior probability distribution for the weights, $p(\bm{\epsilon}^{(I)} | \bm{\phi}, \tilde{\sigma}^2, \alpha^{(I)})$, evaluates to
\begin{equation}
    p(\bm{\epsilon}^{(I)} | \bm{\phi}, \tilde{\sigma}^2, \alpha^{(I)}) = \frac{p(\bm{\phi} | \bm{\epsilon}^{(I)}, \tilde{\sigma}^2) \times p(\bm{\epsilon}^{(I)} | \alpha^{(I)})}{p^{(I)}_{ML}},
\end{equation}
with $\bm{\phi}$ representing the vector of $N_{\mathrm{tr}}$ training log-amplitudes over the data set.
The normalization
\begin{equation}
    p^{(I)}_{ML} = p(\bm{\phi} | \tilde{\sigma}^2, \alpha^{(I)}) = \int d \bm{\epsilon}^{(I)} \, p(\bm{\phi} | \bm{\epsilon}^{(I)}, \tilde{\sigma}^2) \times p(\bm{\epsilon}^{(I)} | \alpha^{(I)})
\end{equation}
is known as the \textit{marginal likelihood}.
Importantly, due to the specific modeling choices, both the posterior distribution as well as the (log) marginal likelihood can be obtained in closed form and are found to be Gaussian distributed \cite{Rasmussen_book,Tipping_2004, fletcher2010relevance}.
The variance of the posterior distribution is given by
\begin{equation}
    \bm{\Sigma}^{(I)} = (\mathbf{K}^{(I) \dagger} \mathbf{B} \mathbf{K}^{(I)} + \alpha^{(I)} \mathbb{1})^{-1},
    \label{eq:posterior_variance}
\end{equation}
where $\mathbf{B}$ denotes a diagonal matrix for which the diagonal elements are given by the inverse likelihood variances for all training configurations, $1/\sigma_{\mathbf{x}}^2$.
The kernel matrix $\mathbf{K}^{(I)}$ consists of the elements $K^{(I)}_{\mathbf{x}, (x', d)} = \tilde{k}^{(I)}_{(x', d)}(\mathbf{x})$, i.e. the $N_{\mathrm{tr}} \times (M D)$ different kernel functions evaluated for all configurations of the training set with the second dimension indexing all compound indices $(x', d)$.
The mean of the posterior, thus also defining the most probable weights according to the regularized modeling assumptions, is given by
\begin{equation}
    \bm{\mu}^{(I)} = \bm{\Sigma}^{(I)} \mathbf{K}^{(I) \dagger} \mathbf{B} \bm{\phi}.
    \label{eq:mean_weights}
\end{equation}
These define the optimal learned weight parameters for site $I$, which depend on the training data as well as two hyperparameters $\tilde{\sigma}^2$ (a global parameter) and $\alpha^{(I)}$ (a site-dependent parameter).

It is key for the regularization and generalization properties of the qGPS model to optimize suitable hyperparameters $\alpha^{(I)}$ and $\tilde{\sigma}$ over the course of the training, for which we follow a \textit{type-II maximum likelihood} approach \cite{Tipping_2004}.
This is based on the maximization of the marginal likelihood, $p^{(I)}_{ML}$, with respect to the hyperparameters.
Maximization of the marginal likelihood is known to find good trade-offs between small training errors and the level of regularization which is applied in order to also generalize the qGPS well outside the training data \cite{Rasmussen_book, Tipping_2004}.
Updates to the quantities of interest can be computed faster for changes in $\alpha^{(I)}$ than for adjustments of $\tilde{\sigma}^2$ because the matrices $\mathbf{K}^{(I) \dagger} \mathbf{B} \mathbf{K}^{(I)}$ do not need to be recomputed.
In our approach, we therefore consider the (log) marginal likelihood optimization separately for $\alpha^{(I)}$ and $\tilde{\sigma}^2$.

In particular, at each fitting step for a selected reference site in the sweeping, we first converge the site-specific value of $\alpha^{(I)}$ via maximizing the (log) marginal likelihood, with the update equation,
\begin{equation}
    \label{eq:alpha_update}
    \alpha^{(I)} \rightarrow \frac{\sum_i (1 - \alpha^{(I)} \bm{\Sigma}^{(I)}_{i,i})}{|\bm{\mu}^{(I)}|^2}.
\end{equation}
This form is based on the update equations used in the context of relevance vector machines \cite{Tipping2000,fletcher2010relevance}.
After the converged value of $\alpha^{(I)}$ is found for the current reference site, we also update the global (site-independent) hyperparameter $\tilde{\sigma}^2$.
This update for $\tilde{\sigma}^2$ is performed by applying a single gradient ascent step with a fixed step size, $\eta$, to Eq.~\ref{eq:marg_like}, to optimize the logarithm of this `noise' hyperparameter, $\ln (\tilde{\sigma}^2)$.
The update of $\tilde{\sigma}^2$ can therefore be expressed as
\begin{equation}
    \label{eq:noise_update}
    \tilde{\sigma}^2 \rightarrow e^{\ln (\tilde{\sigma}^2) + \eta \frac{\mathrm{d} p^{(I)}_{ML}}{\mathrm{d} \tilde{\sigma}^2} \tilde{\sigma}^2}.
\end{equation}
The explicit form for the marginal likelihood as well as its derivative with respect to $\tilde{\sigma}^2$ is given in section \ref{sec:bayesian_appendix} of the appendix where more details are provided.

After the optimal weights are found in closed form for the current site via Eq.~\ref{eq:mean_weights}, and the noise parameter is updated with a single step as described above, we advance to the next site in the lattice where the Bayesian inference and optimization is repeated, starting from the updated value of $\tilde{\sigma}^2$.
The overall optimization algorithm comprises iterative sweeps across the lattice where each sweep infers the optimal local qGPS parameters together with a marginal likelihood maximization at each site of the lattice for appropriate regularization hyperparameters.
In order to track the convergence of the overall scheme, we define the mean log marginal likelihood for a sweep across the lattice as
\begin{equation}
    \lambda = \frac{1}{L} \sum_{I=1}^L  \ln (p^{(I)}_{ML}).
\end{equation}
Here $p^{(I)}_{ML}$ denotes the log marginal likelihood obtained for the Bayesian inference at site $I$ evaluated after optimization of the parameter $\alpha^{(I)}$.
Sweeps across the lattice are then iterated until convergence of the mean log marginal likelihood, $\lambda$, is observed, at which point the qGPS has been fully trained on the given data.
Due to the Bayesian inference on a site-by-site basis, this algorithm does not require any additional (cross) validation and all available training data can directly be used to fit the model.

At convergence of this algorithm, summarized in Algorithm~\ref{alg:sweeping}, we have an optimized posterior distribution for the weights of each site (which fully defines the qGPS), as well as an optimized single, global noise hyperparameter $\tilde{\sigma}^2$, and a set of site-specific hyperparameters, $\alpha^{(I)}$, which define the inverse variance of the prior for the weights on each site $I$.
It is reasonable that the variance of the likelihood of the model should be independent of site, while the $\alpha^{(I)}$ parameters are not, since they control the ease at which the weights on a site can fluctuate.
Its optimal value therefore depends on the kernel function defining the environment of the site, and should therefore be allowed to vary with site.
}

The explicit statistical perspective \vthree{of this sweeping algorithm} contrasts to previous approaches to supervised learning of quantum states which generally involve simple minimization of scalar-valued loss functions, such as the \vone{mean} squared error to fit the model to the (log)-amplitudes (equivalent to a uniform prior), with simple heuristics such as early-stopping to avoid overfitting.
The proposed regularization avoids overfitting to the chosen training data and improves generalization ability, with the marginal likelihood of the model having previously been shown to be an excellent proxy for regularizing the fit of quantum states \cite{PhysRevX.10.041026, doi:10.1063/5.0024570}.
\vthree{The perspective of the sequential, closed-form optimization of local parameters, in the presence of the environmental features defined by the kernel, bears striking resemblance to the density matrix renormalization group algorithm for matrix product states and we expect further synergies between these approaches to be able to be exploited.}

\begin{algorithm}[H]
    \begin{algorithmic}
        \Repeat
        \ForAll{sites $I$}
        \State Set up kernel matrix $\mathbf{K}^{(I)}$ for site $I$ [Eq.~\eqref{eq:quantum_kernel_definition}]
        \State Maximize $\ln (p^{(I)}_{ML})$ w.r.t. $\alpha^{(I)}$ [Eq.~\eqref{eq:alpha_update}]
        \State Update $\epsilon_{d, x', I}$ with mean of posterior, $\bm{\mu}^{(I)}$ [Eq.~\eqref{eq:mean_weights}]
        \State Update $\tilde{\sigma}^2$ based on gradient ascent step [Eq.~\eqref{eq:noise_update}]
        \EndFor
        \Until{$\lambda = \sum_{I=1}^L \ln (p^{(I)}_{ML})$ converges}
    \end{algorithmic}
    \caption{A Bayesian sweeping algorithm for supervised learning with qGPS}
    \label{alg:sweeping}
\end{algorithm}

To demonstrate the ability to overcome the generalization issues for frustrated sign-structures, we compare the Bayesian-regularized sweep fit of the qGPS model to a simple least squares fit in a supervised learning context.
Using the same system and similar numerical setup to Ref.~\onlinecite{Westerhout_2020}, we consider a 2D 24-site Heisenberg model where exact ground-state amplitude training data is selected as a small random sample of \vthree{some small fraction of} the space of all configurations.
The full qGPS wave function model is then trained either via a simple least squares fit to this training data, regularized by keeping $20\%$ of the data as a validation set to determine stopping criteria, or via the Bayesian approach, requiring no separate validation set to minimize the out-of-sample error.
This latter approach instead regularizes the fit through hyperparameters which maximize the marginal likelihood of the model at each step.
This balances the minimization of the in-sample error while also describing the target state well outside the training configurations.
We also compare to an optimization of the qGPS over the entire space of configurations, which illustrates the overall expressibility of the model.

\begin{figure*}[t!]
    \centering
    \begin{subfigure}[b]{0.49\textwidth}
        \includegraphics[width=\textwidth]{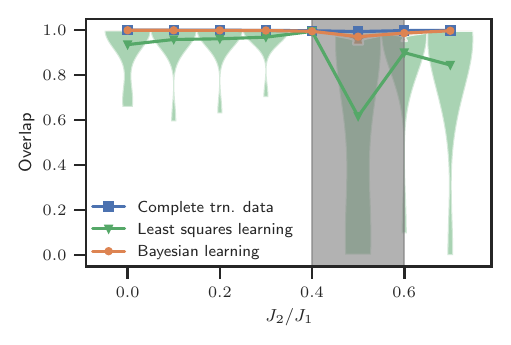}
        \caption{\vthree{Supervised learning using a qGPS with $M=5$ for different values of $J_2/J_1$ using $\sim 1 \%$ of the Hilbert space as training data.
        The grey region indicates the frustrated parameter regime of the system.}}
        \label{fig:supervised_vs_J2}
    \end{subfigure}
    \begin{subfigure}[b]{0.49\textwidth}
        \includegraphics[width=\textwidth]{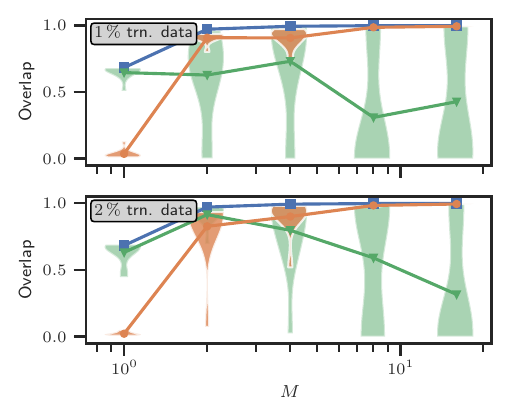}
        \caption{\vthree{Supervised learning using a qGPS with different values of $M$ at the point of high frustration ($J_2/J_1=0.5$). Top (bottom) panel corresponds to a training set of $\sim 1 \%$ ($\sim 2 \%$) of the full Hilbert space size.}}
        \label{fig:supervised_vs_M}
    \end{subfigure}
    \caption{Overlap between the true ground state and a qGPS model obtained by fitting to the ground state of the 2D $J_1$-$J_2$ model on a $4 \times 6$ square lattice.
    Blue squares represent the `ideal' fit, where the \vone{mean} squared error is minimized with respect to the all $\sim 2.7 \times 10^{6}$ configurations of the target state.
    The other two curves show the values obtained by fitting to a restricted subset comprising just $\sim 1 \%$ \vthree{or $\sim 2 \%$} of all ground-state amplitudes, randomly and uniformly sampled from the full Hilbert space.
    The results are shown from the Bayesian learning scheme (orange circles), as well as a minimization of the \vone{mean} squared error over the training data with ADAM and standard validation approaches \cite{DBLP:journals/corr/KingmaB14, bourgin_numpy-ml_2021} (green triangles).
    Both training approaches were repeated with 10 different random training set selections and indicate the mean overlap of the fit qGPS over these training sets.
    The shaded areas of the violin plots indicate the density and spread of optimized fit results across the different training selections.
    Technical details on the training are given in section \ref{sec:bayesian_appendix} of the appendix.}
    \label{fig:overlap_state_fitting}
\end{figure*}

In Fig.~\ref{fig:supervised_vs_J2}, a fit of the qGPS state with $M=5$ on the entire Hilbert space (complete training data) demonstrates excellent expressibility of the model for all levels of frustration.
However, also key is the ability to faithfully describe the state over small, randomly chosen training samples in the frustrated regime, using the Bayesian regularization together with the maximization of the marginal likelihoods (results averaged over independent training samples).
\vthree{For the reported setup, in which training is performed on either $1 \%$ or $2 \%$ of the full set of configurations,} this approach shows a dramatic improvement in the ability to represent the global state from incomplete configurational samples compared to a direct minimization of the least squares error which is regularized by early-stopping based on a validation set error. These latter results are qualitatively similar to Ref.~\cite{Westerhout_2020}, where a similar approach is taken to show regularization errors in the sign structure fitting of NQS, with a particular failure at the maximally frustrated point of $J_2/J_1=0.5$.
Of particular note is the fact that the sweeping approach with Bayesian regularization is significantly less dependent on the specific training set, with only a small scatter of overlaps with the true ground state based on the different training sets, even in the highly frustrated regime.

\vthree{Fig.~\ref{fig:supervised_vs_M} shows the same learning setup, but focussed just at this maximally frustrated point at $J_2/J_1 = 0.5$ and considering increasing qGPS mode complexity as the support dimension, $M$, is increased.
It is expected that with increasing support dimension, the importance of the regularization becomes more significant because the greater model expressibility can lead to overfitting of the limited training data more easily.
The data obtained with a simple minimization of the squared error on the training set, only regularized by early stopping, indeed results in a broad spread of obtained model qualities in the limit of larger support dimensions at deterioration of the result for both training set sizes.
While the data points are in reasonable agreement with the overall model expressivity for the smallest considered support dimension ($M = 1$), the results are significantly less consistent when $M$ is increased.
For both training set sizes of $27042$ (top) and $54083$ (bottom) configurations, corresponding to $\sim 1 \%$ and $\sim 2 \%$ of the full space of configurations, the direct minimization approach mostly fails to reproduce the target state well in this limit of the most complex model.
On the other hand, with the Bayesian sweeping the learned description reliably converges close to the limit of the model expressibility, with the spread across different random realizations even decreasing, as $M$ increased, directly demonstrating that the fit is appropriately regularized even for descriptions of these complex sign structures.
Although the direct minimization of the squared error gives slightly better results for $M=1$, and for the larger training set also at $M=2$, in which case the limited model complexity automatically provides sufficient regularization, the presented results indicate a clear advantage of the supervised learning of more expressive models with the presented Bayesian sweeping approach.
With this demonstrated statistical approach with built-in principled regularization for the modelling of a quantum state expected to represent a significant benefit to the qGPS formulation, we envisage this as an important tool for inferring wavefunctions from limited accessible data. Furthermore, given the established connections of this generalization problem to the optimization difficulties identified with representing complex global sign structures via variational Monte Carlo with from expressive ansatzes \cite{Westerhout_2020}, in the future we aim to combine the Bayesian approach with variational optimization, for an efficient approach for the learning of unknown states \cite{kochkov2018variational, kochkov2021learning}.
}

\section{Conclusion}
For the systems in this work, the qGPS demonstrates excellent overall expressibility and accuracy.
However perhaps more importantly, it also offers tantalizing evidence that it can overcome the numerical and practical bottlenecks of these ML-inspired highly flexible quantum states, by demonstrating that its Bayesian formulation can efficiently regularize the global states for practical optimization in the inevitable case of optimizing the global state from data on highly restricted samples.
This Bayesian regularization and DMRG-inspired algorithm must now be extended to the optimization of an unknown quantum state \cite{kochkov2018variational, kochkov2021learning}, as well as finding application in quantum state tomography \cite{Torlai_2018}.
These new connections between Bayesian ML and quantum many-body systems enabled by the simple qGPS model offer a clear alternative route to bring these fields together, demonstrating state-of-the-art accuracy in challenging many-body problems \vtwo{with further extensions to more general Hamiltonians and Fermionic systems underway \cite{PhysRevB.96.205152, Choo_2020}}.

\begin{acknowledgments}

The authors are thankful for valuable discussions with Aldo Glielmo, Andrew Green and G\'{a}bor Cs\'{a}nyi relating to this work.
G.H.B. gratefully acknowledges support from the Royal Society via a University Research Fellowship, and funding from the Air Force Office of Scientific Research via grant number FA9550-18-1-0515.
The project has also received funding from the European Union's Horizon 2020 research and innovation programme under grant agreement No. 759063.
We are grateful to the UK Materials and Molecular Modelling Hub for computational resources, which is partially funded by EPSRC (EP/P020194/1 and EP/T022213/1).
\end{acknowledgments}

\newpage
\appendix

\section{Representing a qGPS as a Neural Network}
\label{sec:nn_representation_qGPS}
The unsymmetrized qGPS with functional form as presented in Eq.~\eqref{eq:QGPS_def} of the main text associates an amplitude to a many-body configuration $\mathbf{x}$ according to
\begin{equation}
    \Psi(\mathbf{x}) = \exp \left(\sum_{x'=1}^M \prod_{i=1}^L \epsilon_{\mathbf{x}_i, x', i} \right).
\end{equation}
For spin-1/2 systems, where the local Hilbert space dimension is $D=2$ and local occupancies take values $d \in \{\uparrow, \downarrow\}$, this expression can easily be brought into an equivalent form which resembles a specific architecture of feed forward deep neural network model with four layers. \vtwo{This can be useful from both a conceptual framework, in order to see constructive equivalences between classes of states and different perspectives on (for example) how entanglement can efficiently emerge, as well as a practical utility in implementing the qGPS in codebases designed for neural networks. The precise form of this neural network architecture is }given by
\begin{align}
    \Psi(\mathbf{x}) = \exp \Bigg ( \sum_{x'=1}^M \exp \bigg( \sum_{i=1}^L \log \Big(&\epsilon_{\uparrow, x', i} \gamma(\tilde{\mathbf{x}}_i) \nonumber \\ & + \epsilon_{\downarrow, x', i} \gamma(-\tilde{\mathbf{x}}_i)  \Big) \bigg) \Bigg).
\end{align}
The values $\tilde{\mathbf{x}}_i$ correspond to the visible input layer of the neural network, where $\uparrow$  and $\downarrow$ local states on site $i$ are associated with the variables $1$ and $-1$ respectively. The $\gamma$ denotes the activation function associated with the first layer, which takes the form of a rectified linear unit (ReLU), as $\gamma(\tilde{\mathbf{x}}_i) = \text{max}(0, \tilde{\mathbf{x}}_i)$.

Overall, the neural network representation of the qGPS comprises four feed-forward layers where not all layers are fully connected.
The input neurons $\tilde{\mathbf{x}}_i$ represent the test configuration as it is usually done the context of neural quantum states, i.e. they encode the local occupancy as a single number proportional to the corresponding ${\hat S}_z$ eigenvalue of the local occupancy.
A key difference in perspective between the qGPS and a NQS is that the site occupations enter as explicit variables in the NQS, while in a qGPS they enter as simple indices to different variational parameters (similar to how the local site information enters a matrix product state).
The first layer in the above construction ensures that we can transform between these perspectives, allowing the input variables of the NQS representation to be used as an index into the variational parameters.
To do this, it requires $2 \times L$ neurons in the first layer with the ReLU activation.
Indexing the neurons of the first layer by pairs $(j, k)$ with $j = 1 \ldots L$ and $k\in \{\uparrow, \downarrow\}$, the weights connecting the visible units with the first layer are therefore given by $\delta_{i,j} \cdot (\delta_{k,\uparrow} - \delta_{k,\downarrow})$.

This first layer feeds into the second layer of $M \times L$ neurons (indexed by support index $x'$ and site index $i$) with this layer having an activation function of the logarithm, and weights being the variational parameters, $\epsilon_{i, x', k}$, if both neurons are associated with the same site index, or zero otherwise.
The third layer is composed of $M$ neurons, with the weights between second and third layer set to one if both connecting neurons refer to the same support index and to zero otherwise. This is combined with an exponential activation function.
The final output layer all have the same weight of one, with another exponential as the final activation function.

\section{Symmetrization of qGPS}
\label{sec:qGPS_symmetrization}
In practice, it is often helpful to directly include symmetries of the system into the ansatz in order to improve the accuracy of the description and the reliability of the method.
There are two approaches to this symmetrization, which have also both been investigated in the context of NQS.
If the target state approximated by the qGPS is fully symmetric with respect to symmetry operations (i.e. it corresponds to the trivial representation), then we can simply adapt the kernel-symmetrization procedure as it was also applied for the classical GPS \cite{doi:10.1063/5.0024570}.
This is achieved by replacing the kernel function in the model, as defined in section A, by a sum over the kernel functions with respect to all configurations that are symmetrically equivalent to the test configuration.
This `kernel-symmetrized' qGPS, evaluated for a particular test configuration $\mathbf{x}$, is therefore given by the model
\begin{equation}
    \label{eq:QGPS_kernel_symmetrized_def}
    \Psi(\mathbf{x}) = e^{\sum_{x'} \sum_{\mathcal{S}} \prod_i \epsilon_{\mathcal{S}[\mathbf{x}]_i, x', i}}.
\end{equation}
The inner sum includes all symmetry operations $\mathcal{S}$ that we are symmetrizing with respect to, and $\mathcal{S}[\mathbf{x}]$ is the test configuration $\mathbf{x}$ transformed according to these symmetry operations.
Unless stated otherwise, we always consider the qGPS to be symmetrized according to this kernel-symmetrization, effectively corresponding to a product of the ansatz over all symmetrically equivalent copies of the test configuration.
\vthree{If the symmetry operations include all translations across the lattice, as is the case in the setup considered here, then the kernel-symmetrized model can be rationalized in a similar fashion to the construction of NQS with convolutional filters in the network architectures. This effectively ensures the same correlation features at each site of the lattice.
The kernel-symmetrization of the model can also directly be used for the Bayesian sweeping for supervised learning of the state from data as introduced in section \ref{sec:BayesianOpt} of the main text.
In each single step of the sweep, in which the data is fitted with a fixed reference site, all the configurational occupancies are then taken into account at once and the sweeping across the lattice effectively corresponds to a sweep through symmetrized correlation plaquettes of different length scales evaluated for all positions of the lattice at the same time.
}

In addition to the kernel-symmetrization scheme, we can also apply a projective symmetrization approach to symmetrize our representation.
This has recently been shown to significantly help with the optimization of NQS ansatz in order to better capture sign information of the state \cite{Nomura_2021, nomura2021diractype}.
Rather than symmetrizing the kernel function in our ansatz (equivalent to taking a product over all symmetrically equivalent copies of the test configuration), the projective approach applies a sum over the non-symmetrized qGPS amplitudes for the symmetrically equivalent configurations, according to
\begin{equation}
    \label{eq:QGPS_proj_symmetrized_def}
    \Psi(\mathbf{x}) = \sum_{\mathcal{S}}  e^{\sum_{x'} \prod_i \epsilon_{\mathcal{S}[\mathbf{x}]_i, x', i}}.
\end{equation}
In all the results presented in this work, we consider qGPS ansatzes symmetrized according to one of the two approaches.
We always include translational symmetries, the point group symmetries of the lattice, as well as the spin inversion symmetry into the considered set of symmetry operations.

\begin{figure}
    \centering
    \includegraphics[width=\columnwidth]{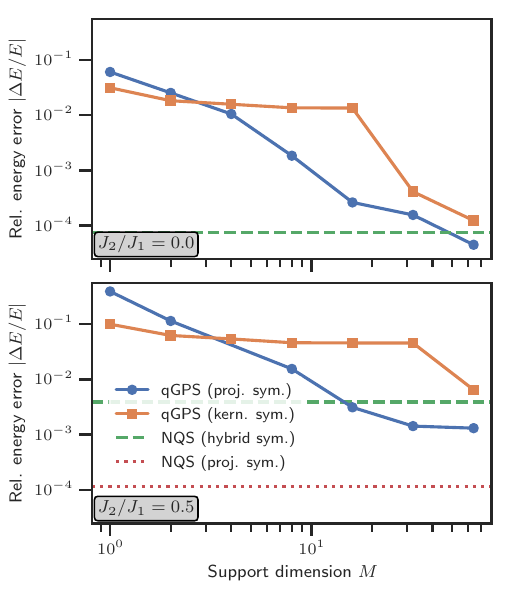}
    \caption{Accuracies achieved with different approaches to symmetrization of the qGPS \vthree{ansatzes} (without Marshall sign rule imposed) for the ground state of two-dimensional spin-1/2 $6\times6$ $J_1$-$J_2$ system, as a function of the support dimension $M$.
    Displayed values correspond to relative energy errors with respect to results obtained with exact diagonalization \cite{refId0}.
    The top figure shows the results for the standard Heisenberg system ($J_2/J_1 = 0.0$) and the bottom shows the $J_2/J_1 = 0.5$ system.
    Blue circles correspond to a kernel symmetrized qGPS ansatz (Eq.~\ref{eq:QGPS_kernel_symmetrized_def}) with orange squares indicating the projectively symmetrized qGPS ansatz (Eq.~\ref{eq:QGPS_proj_symmetrized_def}).
    Reference literature NQS results for comparison are shown with horizontal green dashed lines for Ref.~\onlinecite{PhysRevB.100.125124} and red dotted line for the best NQS results for this system from Ref.~\onlinecite{Nomura_2021}.}
    \label{fig:2d_j1_j2_bothsym}
\end{figure}

If no sign information needs to be captured by the model, we found that the kernel symmetrization approach generally gives better results than the projective symmetrization.
The ansatz then gives, for the studied Heisenberg model, competitive accuracies, even in the limit of rather small support dimension $M$ (see Fig.~\ref{fig:1d_heisenberg} and Fig.~\ref{fig:2d_heisenberg} of the main text).
However, if the model also needs to model sign information, e.g. if the Marshall sign rule is not imposed or in frustrated models where the exact sign structure is not known, we achieved the overall best results for the studied $6 \times 6$  $J_1$-$J_2$ square lattice model using the projectively-symmetrized ansatz.
This observation is exemplified in Fig.~\ref{fig:2d_j1_j2_bothsym}, reporting the relative ground state energy errors achieved with both symmetrization schemes for this setup at $J_2/J_1 = 0$ and $J_2/J_1 = 0.5$ against the support dimension $M$.

In the unfrustrated limit at $J_2/J_1 = 0$, the overall energy errors obtained with the kernel symmetrized approach (indicated by orange squares), mostly fluctuate around approximately $\sim 2 \times 10^{-2}$.
This is significantly worse than the errors presented for the examples where the exact sign structure was imposed explicitly, although the Marshall sign rule can (at least theoretically) be represented \vthree{for this model} (up to a global phase) exactly by a qGPS with support dimension $M=1$.
Although, a sharp drop of the relative energy error to $\sim 4 \times 10^{-4}$ can be observed for the ansatz with $M=32$ and $M=64$ in the figure, it therefore appears that the results are partially influenced by issues with the optimization of the state (rather than only by its expressibility).

In the limit of small support dimensions, $M=1$ and $M=2$, the projectively symmetrized ansatz (datapoints indicated by blue circles in the figure) give slightly worse results than the kernel symmetrization.
However, for larger support dimensions, this scheme yields a much more systematic improvement of the description, with relative energy errors decreasing to almost $10^{-4}$ for $M=64$.
A similar relationship between the performance of the kernel symmetrization as compared to the projective symmetrization can be seen in the bottom part of the figure, showing the results for the frustrated regime at $J_2/J_1 = 0.5$.
While the description accuracy obtained with kernel-symmetrization of the ansatz overall barely increases with increasing bond dimension (apart from a steep increase for the model with $M=64$), the curve for the projectively symmetrized state is much clearer, giving the variationally lower energies between the two approaches for all $M > 4$.
Nonetheless, it is not entirely clear from the data to what extent the obtained results for $J_2/J_1 = 0.5$ could be dominated by optimization issues rather than shortcomings of the ansatz functional form.

The observed relative performance of the two different symmetrization schemes is in keeping with the literature on NQS representations for these systems.
Directly symmetrizing the network in the NQS before applying the final exponentiation (equivalent to our kernel symmetrization scheme), overall gives improved accuracy in regimes where the sign structure can either be directly imposed \cite{PhysRevB.100.125124} or learned efficiently \cite{PhysRevResearch.2.033075}.
However, near the point of maximum frustration at around $J_2/J_1 \sim 0.5$, a loss of accuracy was observed for such NQS representations which have also been attributed to problems with the optimization of the state \cite{10.21468/SciPostPhys.10.6.147, Westerhout_2020}.
Recently a fully projective symmetrization approach was also applied to NQS descriptions which helped overcoming the optimization issues and significantly improved the achieved level of accuracy in the frustrated regime \cite{Nomura_2021, nomura2021diractype, roth2021group}.

In the context of the qGPS, it can be seen that the preferred symmetrization scheme seems to depend on the studied system and choice of support dimension.
Ultimately it would be desirable to define an optimal choice of symmetrization which can be reliably applied to all systems of interest.
In order to introduce such a generally applicable scheme in the future, it will also be crucial to understand how these two symmetrization approaches scale up to larger systems.
Investigations on the one-dimensional Heisenberg model show that, at least in the limit of small support dimensions, the kernel symmetrization has a more consistent level of accuracy as the system grows in size (as shown in Fig.~\ref{fig:1d_heisenberg} of the main text), and therefore might be better suited for the study of larger systems. However, combinations of the two approaches, as have previously been used in NQS studies, may also be advantageous.

\section{Simulation details in Variational optimization}
\label{sec:VMC_details}
All results presented in this work were obtained using the NetKet software package \cite{CARLEO2019100311, https://doi.org/10.48550/arxiv.2112.10526}.
Where the qGPS is directly and variationally optimized, the parameters of the model were updated using the standard Stochastic Reconfiguration (SR) approach \cite{PhysRevB.64.024512}.
In the SR, the variational parameters of the model are updated at each optimization step according to
\begin{equation}
    \label{eq:parameter_updates}
    \mathbf{p}^{(k)} = \mathbf{p}^{(k-1)} - \tau \, (\mathbf{S}^{-1} \cdot \textbf{grad}(E)),
\end{equation}
where $\mathbf{p}^{(k)}$ is the vector of parameters after the $k$-th parameter update, $\tau$ defines a step size, and $\textbf{grad}(E)$ denotes the gradient of the variational energy with respect to the parameters.
The overlap matrix, $\mathbf{S}$, whose inverse is multiplied with the energy gradient in the update equation can be understood as a preconditioner for a gradient descent based minimization of the variational energy and typically helps with the convergence of the optimization algorithm.
As it is common practice, we add a constant shift, $c$, to the diagonal elements of $\mathbf{S}$ in all our VMC calculations in order to stabilize the inversion of the overlap matrix.

The overlap matrix and the energy gradient can be cast in terms of expectation values which can be stochastically sampled from batches of configurations with Markov chain Monte Carlo sampling.
To achieve this, operators $\hat{O}_j$, which encode the derivatives of the log-wavefunction amplitudes with respect to the variational parameter $p_j$, are introduced.
More specifically, these log-amplitude derivatives are the eigenvalues of the introduced operators $\hat{O}_j$ which are diagonal in the computational basis and are formally defined as
\begin{equation}
    \langle \mathbf{x} | \hat{O}_j | \mathbf{y} \rangle = \delta_{\mathbf{x},\mathbf{y}} \frac{\mathrm{d} \ln(\Psi (\mathbf{x}))}{\mathrm{d} p_j}.
\end{equation}
The element of the energy gradient corresponding to parameter $p_j$ then evaluates to
\begin{equation}
    \text{grad}_j(E) = \langle \hat{O}_j^\ast \hat{H} \rangle - \langle \hat{H} \rangle \langle \hat{O}_j^\ast \rangle,
\end{equation}
and the elements of the overlap matrix are given by the expression
\begin{equation}
    S_{j,k} = \langle \hat{O}_j^\ast \hat{O}_k \rangle - \langle \hat{O}_j^\ast \rangle \langle \hat{O}_k \rangle.
\end{equation}

For the VMC results presented in this work, the final qGPS representation was obtained by iteratively applying the updates according to Eq.~\eqref{eq:parameter_updates} to the parameters of the model.
We chose a default update step size of $\tau = 0.02$.
Although it is in principle possible to use real parameters if no sign information needs to be described by the model, we always chose the parameters to be complex valued.
Expectation values were evaluated via standard Markov chain Monte Carlo sampling where the number of sampled configurations was increased during the optimization.
Presented results always refer to a final evaluation of the variational energy (with an increased number of samples) for the final qGPS model.
The final qGPS parameters were taken to be those which resulted in the smallest sum of the sampled variational energy and estimated error from the last $40$ optimization steps.

Since parameter updates can sometimes result in numerical issues such as overflows or instabilities, we always revert to the previous parameter values if invalid values (such as overflows) are observed in the sampling or in the evaluation of the expectation values after an update.
The update step size is then decreased and the diagonal shift increased by a factor of two, before the parameter updates are recomputed with these adjusted values.
This adjustment of the parameters is iterated until a set of numerically feasible parameters is found in the update at which point the step size and diagonal shift are reset back to their default values.
Nonetheless, we found that numerical instabilities and sub-optimal optimization trajectories might still be observed in the optimization of the ansatz depending on the initialization of the parameters as well as the specifics of the algorithm.
Similar numerical instabilities in the optimization also appear to be common when optimizing NQS architectures \cite{PhysRevResearch.2.033075, 10.21468/SciPostPhys.10.6.147}.
Devising general, universally applicable methods and approaches to optimize such highly expressive model in a stable and reliable way, could and should be a key element for future research.
In the following, we present the numerical details of our VMC calculations which were specific to the considered systems.

The results for the 1D Heisenberg model presented in Fig.~\ref{fig:1d_heisenberg} of the main text were obtained with incorporation of the Marshall sign rule into the Hamiltonian (i.e. no sign information needs to be described) \cite{doi:10.1098/rspa.1955.0200, carleo2018constructing, PhysRevResearch.2.033075}.
The product state results were obtained using a single product state parametrization which was symmetrized with respect to the same symmetries as the qGPS, by summing together the amplitudes of all symmetrically equivalent configurations.
This ansatz therefore corresponds to a kernel-symmetrized qGPS ansatz with $M=1$ where the final exponentiation is not applied (effectively corresponding to a projective symmetrization of the state).
The parameters $\epsilon_{d, x', i}$ were initialized as random phase factors with phases normally distributed about zero, with standard deviation of $0.02$ and unit magnitude.
Also shown are results from a two-body Jastrow ansatz as implemented in Netket (with the same symmetries imposed as for the other ansatzes \vthree{except for the spin inversion symmetry}), for which we applied the same optimization protocol as for qGPS and product state ansatz.
Figure~\ref{fig:2d_heisenberg} of the main text shows the results for the 2D $10\times10$ Heisenberg model ground state with the kernel-symmetrized qGPS.
Again, the problem was transformed to incorporate the Marshall sign rule so that the qGPS model only needs to describe the magnitude of the wave function and not its sign structure.
The specific details of the VMC simulation are mostly the same as for the one-dimensional setting.
The parameters were again initialized as random phase factors were the associated phases are drawn from a normal distribution located at zero.
We chose the value of the standard deviation as $0.02$ for bond dimensions smaller than or equal to $M=10$ and as $0.01$ otherwise.

The data points presented in Figs.~\ref{fig:1d_heisenberg} and \ref{fig:2d_heisenberg} of the main text correspond to calculations where the amplitudes of the modelled target state all have the same sign.
For calculations on the two-dimensional square lattice of $6 \times 6$ sites, we did however not employ the transformation encoding the Marshall sign rule, either in the frustrated or unfrustrated points.
Therefore, the amplitudes of the modelled target state comprised a sign structure for both considered values of $J_2$.
The obtained results are shown in Fig.~\ref{fig:2d_j1_j2} in the main text as well as in Fig.~\ref{fig:2d_j1_j2_bothsym} in appendix \ref{sec:qGPS_symmetrization}.
The figure in the main text only shows the results obtained with the projectively symmetrized ansatz whereas the appendix section also includes the results obtained for this system with the kernel-symmetrized model.

The parameters of the ansatzes were again initialized with random phase factors.
The phases were drawn from a normal distribution around zero with a standard deviation of $0.05$ for the projectively symmetrized ansatzes and with a standard deviation of $0.02$ for the kernel symmetrized ansatz.
Only the magnitudes of the initial parameters associated with one site were for the kernel symmetrized ansatz not chosen to be equal to one.
The magnitudes of these initial parameters were drawn from the same normal distribution as the phases.
The more expressive ansatzes for this frustrated system with support dimension $M=64$ were optimized with $3000$ optimization steps using an initial number of samples of \vtwo{$\sim$}$10100$.
This number of samples was then increased every $40$ iterations by $100$ up to a total number of \vtwo{$\sim$}$18000$ samples for the last $40$ optimization steps.
For these calculations with the largest support dimension, we also chose an initial default diagonal shift of $c=0.02$ (as compared to a fixed value of $c=0.01$ for calculations on less expressive ansatzes and other systems) which was decreased by a factor of $0.97$ whenever the number of samples was increased.

For the kernel symmetrized ansatz it is possible to only exponentiate ratios of wave function amplitudes, which is intrinsically numerically stable.
For the projectively symmetrized ansatz however, it is necessary to evaluate a sum over exponentiated expressions which requires care for numerical stability.
We therefore also included a term into our model, which rescaled the magnitudes of all qGPS models according to $\Psi(\mathbf{x}) \rightarrow \Psi(\mathbf{x})\times e^{-b}$.
By updating the bias, $-b$, after each optimization step, the overall scale of the amplitudes can be controlled.
This does not change the general expressibility of the model but it can help to avoid numerical issues caused by amplitudes becoming too large.
We set the bias $b$ to the maximum real component of the log-amplitudes sampled from the configurations in the previous update step.
This enforces the magnitudes of the wave function amplitudes to be approximately between zero and one.

\section{Implementation details of the supervised Learning with qGPS}
\label{sec:bayesian_appendix}


In this section, we outline the technical details for the supervised learning with the qGPS for which results are presented in Fig.~\ref{fig:overlap_state_fitting} of the main text.
We considered a setup very similar to the one presented in Ref.~\onlinecite{Westerhout_2020}.
In particular, we studied the task of learning a qGPS representation
from the exact ground state data of two-dimensional $J_1$-$J_2$ models on a $4 \times 6$ site square lattice.
The learning was done by training the model based on the exact wave function data associated with $27,042$ and $54,083$, randomly selected basis configurations, corresponding to $\sim 1\%$ and $\sim 2\%$ of the full Hilbert space size.
We considered two different approaches to achieve this goal.

The first approach considered, conceptually similar to the approaches used for the optimization of the NQS as presented in Ref.~\onlinecite{Westerhout_2020}, is based on standard techniques from the field of machine learning.
It consists of the minimization of the squared error of the amplitudes predicted by the qGPS model $\Psi$ with respect to the target amplitudes $\Psi_{\rm{target}}$ for the configurations in the training set.
This means the parameters of the qGPS are found by minimizing the loss function
\begin{equation}
    \mathcal{L} = \sum_{\{\mathbf{x}\}}^{N_{\mathrm{tr}}} \vert \Psi(\mathbf{x}) - \Psi_{\rm{target}}(\mathbf{x}) \vert ^{2},
\end{equation}
where $\{\mathbf{x}\}$ denotes the set of all training configurations.
The loss function can be minimized using different optimizers, here we considered the ADAM optimization scheme \cite{DBLP:journals/corr/KingmaB14}.
As it is the standard for such learning approaches, we split the training set into multiple small batches in each minimization epoch which were sequentially used to compute the parameter updates.
Further regularization of the fit was achieved by holding back $20\%$ of the training configurations which are used to estimate the error outside of the data used for the fit (out-of-sample error).
This validation set was used to determine at which optimization step the ideal generalization of the model beyond the training data was obtained.
This early stopping regularization can help to prevent overfitting of the training data.
Nonetheless, the overfitting of the training data still emerged as a key problem in this approach, especially for the target state in the frustrated regime where $J_2/J_1 \sim 0.5$, as found in the NQS literature \cite{Westerhout_2020}.

As an alternative to the standard minimization of the squared error, we introduce an alternative approach to learn a qGPS representation from given data.
This approach is very specific to the functional form of the qGPS and it explicitly exploits the fact that it can be reformulated as a form of exponentiated kernel model.
The aim is to learn the variational parameters one site at a time with rigorous Bayesian inference techniques.
The qGPS is then learned in an iterative way by repeatedly sweeping across the lattice, similar to a DMRG optimization.
As this approach is based on iterative Bayesian inference for each site, it can help with learning the qGPS in a robust and reliable way for many different settings.
\vthree{The main concepts of this Bayesian sweeping approach are outlined in the main text of the manuscript and we present the specific implementation details in the following.

Key to the Bayesian learning scheme considered here is that the qGPS model can be rewritten as a kernel model of the form
\begin{equation}
    \Psi(\mathbf{x}) = \exp \left(\sum_{m=1}^{M \times D} \epsilon^{(I)}_m \, \tilde{k}_{m}(\mathbf{x}) \right).
    \label{eq:kernel_form_qGPS}
\end{equation}
The introduced weights $\epsilon^{(I)}_m$ correspond to the parameters $\epsilon_{d, x', I}$ of the original model where the compound index $m = (x', d)$ includes all pairs of indices $d$ and $x'$ for the selected site.
The kernel function, $\tilde{k}_{m}(\mathbf{x})$, is obtained via a symmetrization of the parameters on all other sites, as shown in Eq.~\eqref{eq:quantum_kernel_definition} of the main text and is given by
\begin{equation}
    \tilde{k}_{m}(\mathbf{x}) = \sum_\mathcal{S} \delta_{\mathcal{S}[\mathbf{x}]_I, d} \cdot \prod_{i \neq I}^L \epsilon_{\mathcal{S}[\mathbf{x}]_i, x', i}.
\end{equation}
This kernel function can be interpreted as a comparison between a test configuration $\mathbf{x}$ and an artificial (not directly specified) quantum support configuration labelled by $m$.
Alternatively it can also be understood as a specific renormalized basis representation, mapping the configuration $\mathbf{x}$ to a feature defined by $\tilde{k}_{m}(\mathbf{x})$.

We adapt Bayesian learning techniques to learn the vector of weights, $\bm{\epsilon}^{(I)}$, based on the available log-wavefunction training data (specified by the set of training configurations $\{\mathbf{x}\}$ and the vector containing the associated log-wavefunction amplitudes, $\bm{\phi}$).
This is mostly similar to the approach presented in Ref.~\onlinecite{PhysRevX.10.041026,doi:10.1063/5.0024570}.
Specifically, we model the likelihood of the log-wavefunction data as a normal distribution around the predicted log-amplitude of the qGPS with a data dependent variance, $\sigma^2_{\mathbf{x}}(\tilde{\sigma}^2)$, as specified in Eqs.~\eqref{eq:likelihood} and \eqref{eq:sigma_def} of the main text.
The prior distribution of the weights is also assumed to be normal with zero mean and a variance characterised by an additional per-site hyperparameter $\alpha^{(I)}$.
}
Since the weights and the amplitudes are in this work generally considered to be complex valued, all probability distributions used for the Bayesian inference are probability distributions of complex random variables \cite{halliwell2015complex}.
The descriptions for real random variables can however easily be extended by replacing the probability distributions with specific complex equivalents \cite{RJ-2015-006}.
The central elements of the Bayesian inference for complex-valued random variables as used in this work are outlined in the following.
Note that we use the symbol $^\dagger$ to refer to the hermitian conjugate of a matrix.

\vthree{Under the stated modelling assumptions, both the posterior distribution for the weights $\bm{\epsilon}^{(I)}$ as well as the (log) marginal likelihood obtained with the given training data can be expressed in closed form.
As presented in the main text (Eqs.~\eqref{eq:mean_weights} and \eqref{eq:posterior_variance}), the posterior is a Gaussian distribution with mean
\begin{equation}
    \bm{\mu}^{(I)} = \bm{\Sigma}^{(I)} \mathbf{K}^{(I) \dagger} \mathbf{B} \bm{\phi}
\end{equation}
and variance
\begin{equation}
    \bm{\Sigma}^{(I)} = (\mathbf{K}^{(I) \dagger} \mathbf{B} \mathbf{K}^{(I)} + \alpha^{(I)} \mathbb{1})^{-1}.
\end{equation}
}
The logarithm of the marginal likelihood can be expressed as \cite{fletcher2010relevance, doi:10.1063/5.0024570}
\begin{multline}
    \ln(p^{(I)}_{ML}) = M \times D \times \ln(\alpha^{(I)}) - \sum_{\{\mathbf{x}_{tr}\}} \ln(\pi \sigma_{\mathbf{x}_{tr}}^2) \\
         + \ln(\det(\bm{\Sigma}^{(I)})) - \bm{\phi}^\dagger  \mathbf{B} \bm{\phi} + \bm{\mu}^{(I) \dagger} (\bm{\Sigma}^{(I)})^{-1} \bm{\mu}^{(I)}. \label{eq:marg_like}
\end{multline}

\vthree{The hyperparameters $\alpha^{(I)}$ and $\tilde{\sigma}$ are updated by iteratively maximizing the log marginal likelihood.
We first update the parameter $\alpha^{(I)}$ according to the update equation (Eq.~\eqref{eq:alpha_update} of the main text)}

\begin{equation}
    \alpha^{(I)} \rightarrow \frac{\sum_i (1 - \alpha^{(I)} \bm{\Sigma}^{(I)}_{i,i})}{|\bm{\mu}^{(I)}|^2}.
\end{equation}
This form is based on the update equations used in the context of relevance vector machines \cite{Tipping2000,fletcher2010relevance}.
\vthree{Afterwards, we apply a single gradient ascent step to the logarithm of the parameter $\tilde{\sigma}^2$ resulting in the update (Eq.~\eqref{eq:noise_update} of the main text)}
\begin{equation}
    \tilde{\sigma}^2 \rightarrow e^{\ln (\tilde{\sigma}^2) + \eta \frac{\mathrm{d} p^{(I)}_{ML}}{\mathrm{d} \tilde{\sigma}^2} \tilde{\sigma}^2}.
\end{equation}
The derivative of the marginal likelihood with respect to $\tilde{\sigma}^2$ is given by
\begin{multline}
        \frac{\mathrm{d} p^{(I)}_{ML}}{\mathrm{d} \tilde{\sigma}^2} = \text{tr}(\mathbf{B}' \mathbf{B}^{-1} - \mathbf{K}^{(I) \dagger} \mathbf{B}' \mathbf{K}^{(I)} \bm{\Sigma}^{(I)}) - \bm{\phi}^{\dagger}\mathbf{B}'\bm{\phi} \\ - \bm{\mu}^{(I) \dagger} \mathbf{K}^{(I) \dagger} \mathbf{B}' \mathbf{K}^{(I)} \bm{\mu}^{(I)}  + 2 \bm{\phi}^{\dagger}\mathbf{B}' \mathbf{K} \bm{\mu}^{(I)} ,
\end{multline}
with a diagonal matrix $\mathbf{B}'$ with diagonal elements
\begin{equation}
    \mathbf{B}' = \frac{\mathrm{d} (1/\sigma_{\mathbf{x}_{tr}}^2)}{\mathrm{d}\tilde{\sigma}^2 } =  - \frac{1}{\sigma_{\mathbf{x}_{tr}}^4 ( |\Psi(\mathbf{x})|^2 +\tilde{\sigma}^2)}.
\end{equation}
\vthree{This derivative can be used for the gradient ascent updates to the noise hyperparameter $\tilde{\sigma}^2$ as outlined in the main text.}

\vthree{Fig.~\ref{fig:overlap_state_fitting} of the main} text shows the mean overlap with the target states averaged across different random training sets, obtained with the Bayesian inference described above and a more traditional least squares error minimization with validation.
The plot also contains the overlap with the target states when fitting the model on the data of the complete Hilbert space for reference.
The mean overlaps were calculated by repeating each of the two training schemes with ten different random training sets (using the same random realizations in both approaches).
The figure also visualizes the spread of the obtained overlaps for the different random training sets with violin plots.
In order to set an appropriate overall scale of the training amplitudes, we renormalize them for all of the approaches so that the log-amplitudes of the training set have zero mean \cite{doi:10.1063/5.0024570}.
Moreover, in each approach, we initialized the qGPS with random values for the model parameters $\epsilon_{\mathbf{x}_i, x', i}$.
The real and imaginary parts of the initial parameters were all drawn from Gaussian distributions with zero mean and a standard deviation of $0.5$.

In the simple least squares minimization approach, we took $20 \%$ of the whole training data (picked at random) as a validation set and trained the model by fitting on the remaining $80 \%$ of the training data.
The parameters of the model were updated with the ADAM optimizer using randomly chosen batches of $64$ configurations.
The last batch in each optimization epoch was smaller so that each data point was only considered once in an epoch.
After each epoch, we evaluated the mean squared error for the validation set and stopped the optimization if no improvement in the validation set error was observed over 10,000 successive epochs \vthree{or the optimization was leading to numerical instabilities}.
We used the ADAM optimizer as implemented in the \textit{numpy-ml} package \cite{bourgin_numpy-ml_2021} using two different learning rates of $10^{-3}$ and $10^{-4}$ together with default values for the other optimization parameters.
Those parameters for which the validation error after an epoch was the smallest across all iterations and both learning rates then defined the final model.
\vthree{In some instances, the optimized model could not be evaluated over the full data set due to numerical issues (thus indicating very significant overfitting problems).
In these cases we included a data point with vanishing overlap into the statistics visualized in Fig.~\ref{fig:overlap_state_fitting}.}

In the iterative Bayesian learning approach, we initialized all $\alpha^{(I)}$ values with a value of $2$.
The noise parameter $\tilde{\sigma}^2$, which corresponds to the fixed variance of the likelihood of the inferred amplitudes, was initially set to the mean squared error of the initial model across the training set.
Afterwards, its logarithm, $\ln (\tilde{\sigma}^2)$, was updated after the optimization at each site by applying gradient ascent steps with respect to the log marginal likelihood, with a step size of $\eta = 10^{-5}$.
At early stages of the optimization, $\tilde{\sigma}^2$ can sometimes grow to very large values, where the data is not yet described well by the learned model.
Such a very large noise parameter can prevent the algorithm from appropriately learning from the given data.
We therefore never let the noise parameter, $\tilde{\sigma}^2$, increase beyond its initial value, which should represent an upper bound.
The sweeps across the lattice were done in the same order for every sweep, scanning across the lattice one row after the other.

Finally, the results from fitting on the complete training set were obtained from \vthree{minimizing}
the full \vone{mean} squared error with a quasi Newton optimizer (BFGS).
\vthree{These calculations were intialized by either first running 1,000 epochs with the ADAM minimization scheme on the full data set with a learning rate of $10^{-3}$ and adapting the best parameters out of these 1,000 epochs (if $M \leq 2$), or by first running $10$ Bayesian sweeps w.r.t. the full data set across the lattice as described above (if $M > 2$).}
While the Bayesian sweeping algorithm might be improved by additional adjustments to the specific details of the algorithm in the future, the obtained results already indicate a great potential of this approach for learning appropriate models of quantum states from randomly sampled training data in a reliable, generalizable and robust way.

%

\end{document}